\documentclass[twocolumn, prb, a4paper, 10pt, superscriptaddress, nopacs, floatfix, longbibliography]{revtex4-1}

\usepackage{mathtools}
\usepackage{siunitx}
\usepackage{graphicx}
\usepackage{xcolor}
\usepackage{physics}
\usepackage{nicefrac}
\usepackage{bm}
\usepackage{float}
\usepackage[utf8]{inputenc}
\usepackage[colorlinks, citecolor=blue, linkcolor=black]{hyperref}
\usepackage{ulem}
\usepackage{booktabs}
\usepackage{multirow}
\usepackage{xspace}
\usepackage{array}

\newcommand{\VB}{V$_\text{B}^-$\xspace}

\begin{document}

\title{Optimizing the performances of quantum sensing foils\\
 based on boron vacancy centers in hexagonal boron nitride}

\author{J. Lacaze}
\author{T. Clua-Provost}
\affiliation{Laboratoire Charles Coulomb, Université de Montpellier and CNRS, 34095 Montpellier, France}
\author{T. Poirier}
\affiliation{Tim Taylor Department of Chemical Engineering, Kansas State University, Manhattan, Kansas 66506, USA}
\author{L. Houzelle}
\affiliation{Laboratoire Charles Coulomb, Université de Montpellier and CNRS, 34095 Montpellier, France}
\author{C.~Crunteanu-Stanescu}
\author{C.~Robert}
\affiliation{Universit\'e de Toulouse, INSA-CNRS-UPS, LPCNO, 135 Avenue Rangueil, 31077 Toulouse, France}
\author{T.~Michel}
\author{B.~Gil}
\affiliation{Laboratoire Charles Coulomb, Université de Montpellier and CNRS, 34095 Montpellier, France}
\author{G. Cassabois}
\affiliation{Laboratoire Charles Coulomb, Université de Montpellier and CNRS, 34095 Montpellier, France}
\affiliation{Institut Universitaire de France, 75231 Paris, France}
\author{J.~H.~Edgar}
\affiliation{Tim Taylor Department of Chemical Engineering, Kansas State University, Manhattan, Kansas 66506, USA}
\author{V. Jacques}
\email{vincent.jacques@umontpellier.fr}
\affiliation{Laboratoire Charles Coulomb, Université de Montpellier and CNRS, 34095 Montpellier, France}

\begin{abstract}
Boron vacancy (\VB) centers in hexagonal boron nitride (hBN) offer a promising route toward quantum sensing and imaging technologies based on a two-dimensional material platform. In this work, we investigate the optical and spin properties of \VB center ensembles generated by neutron irradiation of monoisotopic h$^{10}$B$^{15}$N crystals over a broad range of irradiation doses, with the aim of identifying the irradiation conditions that optimize magnetic field sensing performance. By systematically comparing measurements performed on hBN flakes of identical thickness ($\sim 20$~nm), we first identify an irradiation threshold beyond which the photoluminescence intensity decreases, reflecting both a reduction in the quantum efficiency of \VB centers and a saturation of their density at high doses. We then study the dependence of the spin properties on neutron irradiation dose, focusing on the zero-field splitting parameters, the optically detected magnetic resonance (ODMR) response, and spin relaxation dynamics. These measurements identify an optimal neutron irradiation regime yielding a DC magnetic field sensitivity of $\eta_{\rm opt}\sim 10~\mu$T/$\sqrt{\rm Hz}$ under diffraction-limited laser excitation, while maintaining spin coherence times close to the limit imposed by fluctuations of the intrinsic nuclear spin bath in hBN. Our results provide important guidelines for optimizing \VB center ensembles in hBN for two-dimensional quantum sensing applications.
\end{abstract} 

\maketitle

\section{Introduction}

The negatively charged boron vacancy (\VB) center in hexagonal boron nitride (hBN) has recently attracted considerable attention as a two-dimensional (2D) platform for quantum sensing technologies~\cite{ReviewTongCanLi,Vogl2026}. This point defect hosts a spin-triplet ground state that can be efficiently polarized by optical pumping, coherently manipulated with microwave fields, and read out through its spin-dependent photoluminescence (PL)~\cite{Gottscholl2020,Gottscholl2021,haykal2021decoherence}. These properties enable local sensing and imaging of a wide range of physical quantities, including magnetic fields, temperature, pressure, and strain~\cite{Du2021,PhysRevApplied.18.L061002,Tetienne2023,Rita2024,Toeno26,Mu2025,Harvard2025,GaoStrain2022,ZhaoStrain2026}. A key advantage of this sensing platform is the ability to exfoliate thin hBN flakes and deterministically transfer them onto virtually any substrate, enabling straightforward integration of the spin-based sensor with a broad range of material systems. Moreover, reducing the hBN thickness to only a few atomic layers could bring \VB centers within nanometer-scale proximity of a target sample, offering a potential advantage for nanoscale magnetic sensing.

\indent A limitation of the \VB center as a quantum sensor, however, is the low quantum efficiency of its optical transition~\cite{ivady2020,Reimers2020,Jules2026}, which results in very dim emission that precludes single-defect detection. Dense ensembles of \VB centers are therefore essential for achieving high-sensitivity sensing and imaging. Two main strategies can be envisioned for fabricating quantum sensors based on \VB centers in hBN. The first consists of exfoliating thin flakes from a high-purity hBN crystal and subsequently generating \VB centers by ion implantation. In this approach, the ion species, implantation energy, and dose must be carefully optimized for the target thickness of the hBN layer in order to control both the depth distribution and density of the generated defects~\cite{IgorImplant2020,ACSomega2022,HeImplant2023,PhysRevMaterials.9.056203}. The second strategy relies on neutron irradiation of a bulk hBN crystal, which efficiently generates \VB centers across the full crystal volume by nuclear transmutation~\cite{Li2021}. The irradiated material can then be mechanically exfoliated to readily obtain sensing flakes of different thicknesses for direct transfer onto the sample under study. A practical limitation of this approach, however, is the need for a nuclear reactor, a large-scale research facility available at only a limited number of sites. Nevertheless, neutron irradiation remains the only demonstrated route to date for obtaining \VB centers in few-atomic-layer thick hBN flakes~\cite{Jules2026,DurandPRL2023,Rita2024}, making it particularly promising for the development of hBN-based quantum sensing technologies.\\
\indent In this work, we systematically investigate the optical and spin properties of \VB center ensembles generated by neutron irradiation of hBN crystals over a broad range of doses, with the goal of identifying the irradiation conditions that provide the best magnetic field sensing performance. To this end, we use monoisotopic h$^{10}$B$^{15}$N crystals, which provide the most favorable isotopic environment for \VB-based sensing. In this configuration, \VB centers exhibit a simplified hyperfine structure, with narrower magnetic resonances and slightly enhanced spin coherence properties~\cite{Clua2023,Gong2024}. Moreover, $^{10}$B has one of the largest neutron capture cross sections of all isotopes in the periodic table, enabling highly efficient generation of \VB centers by neutron-induced nuclear transmutation~\cite{Li2021}. To enable meaningful comparisons across irradiation doses, all measurements are performed on exfoliated hBN flakes with identical thickness ($\sim20$~nm). We start by studying the optical brightness of these hBN flakes. Our results reveal an irradiation threshold beyond which the PL intensity decreases, reflecting both a reduction in the quantum efficiency of \VB centers and a saturation of their density. These effects are likely associated with the formation of extended defects and vacancy clusters at high irradiation doses. We then analyze how the spin properties of \VB center ensembles evolve with irradiation dose focusing on (i) the zero-field splitting parameters of the spin-triplet ground state, (ii) the optically detected magnetic resonance (ODMR) response and (iii) the spin relaxation properties. From these measurements, we identify a neutron irradiation regime that yields an optimal DC magnetic field sensitivity of about $\eta_{\rm opt}\sim 10~\mu$T/$\sqrt{\rm Hz}$ under diffraction-limited laser excitation while maintaining the spin coherence time close to the limit set by fluctuations of the intrinsic nuclear spin bath in hBN.

\begin{figure}[t]
  \centering
  \includegraphics[width = 8.5cm]{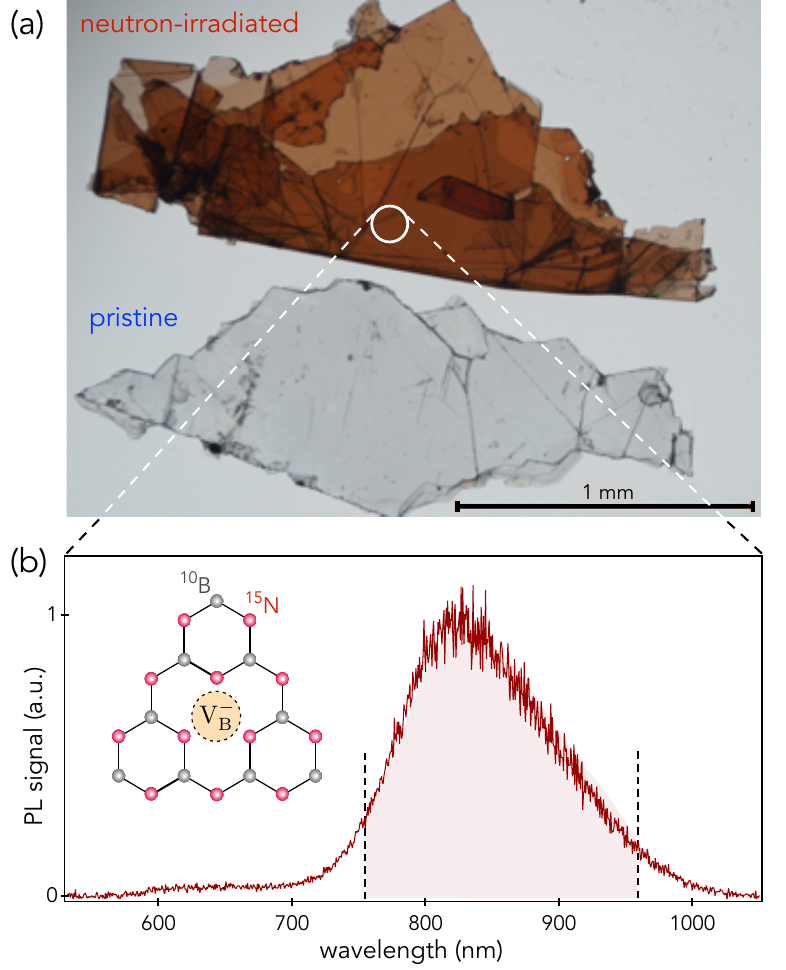}
  \caption{(a) Photograph of a pristine h$^{10}$B$^{15}$N crystal (bottom) grown by the metal-flux method. After irradiation with thermal neutrons, the crystal becomes reddish due to the creation of \VB centers (top). (b) Typical PL spectrum recorded from sample S8 under green laser illumination at room temperature. The red shaded area corresponds to the spectral window detected by the avalanche photodiode of the scanning confocal microscope. The inset shows the microscopic structure of the \VB center in h$^{10}$B$^{15}$N.}
  \label{fig1}
\end{figure}
\section{Materials and methods}

Bulk hBN crystals were synthesized via the metal-flux growth method described in Ref.~\cite{EdgarAdvMat}, using boron powder isotopically enriched in $^{10}$B (99.2\%) and nitrogen gas enriched in $^{15}$N ($>99\%$) as starting materials. The resulting monoisotopic h$^{10}$B$^{15}$N crystals, which have typical lateral dimensions in the millimeter range and thicknesses of a few tens of micrometers [Fig.~1(a)], were irradiated with thermal neutrons at the Ohio State University Research Reactor.  Due to their long mean free path, thermal neutrons generate \VB centers throughout the entire thickness of the hBN crystal, imparting a characteristic reddish color to the material [Fig.~1(a)]. Table 1 summarizes all the samples (S1--S9) studied in this work, with neutron irradiation doses spanning nearly two orders of magnitude, from $1.1\times 10^{16}$neutrons/cm$^{2}$ to $8.7\times 10^{17}$neutrons/cm$^{2}$.

\begin{table}[b]
\centering
\setlength{\extrarowheight}{0.15cm}
\begin{tabular}{cccccccc}
	\hline
	 \hline
 & sample No.  &&&&&&  irradiation dose (neutrons/cm$^{2}$) \vspace{0.1cm} \\
\hline

  & S1 &&&&&& $1.1\times 10^{16}$  \\

  & S2 &&&&&& $2.2\times 10^{16}$  \\

  & S3   &&&&&& $2.9\times 10^{16}$  \\

  & S4 &&&&&&  $3.6\times 10^{16}$ \\
  & S5 &&&&&&  $4.3\times 10^{16}$ \\
    & S6 &&&&&&  $5.8\times 10^{16}$ \\
      & S7 &&&&&&  $7.2\times 10^{16}$ \\
        & S8 &&&&&&  $1.4\times 10^{17}$ \\
        & S9 &&&&&&  $8.7\times 10^{17}$ \\
      \hline 
      \hline
\end{tabular}
\caption{Summary of the neutron-irradiated h$^{10}$B$^{15}$N crystals studied in this work.} 
\label{tab:table}
\end{table} 

For each irradiation dose, hBN flakes were mechanically exfoliated from the corresponding neutron-irradiated crystal and transferred onto a SiO$_2$(85~nm)/Si substrate. The optical and spin properties of \VB center ensembles were studied using a scanning confocal microscope operating under ambient conditions. Laser excitation at $532$~nm was focused onto the hBN flake through a microscope objective with a numerical aperture of $0.65$. The PL signal was collected by the same objective, spectrally filtered to remove the excitation laser, and directed either to a spectrometer or to a silicon avalanche photodiode (APD) operating in the photon-counting regime. A typical PL spectrum recorded from a neutron-irradiated hBN sample is shown in Fig.~1(b). It features the broad near-infrared emission band characteristic of \VB centers~\cite{Gottscholl2020}. For measurements of the PL signal with the APD, we used a combination of a $750$-nm long-pass filter and a $945$-nm short-pass filter [Fig.~1(b)]. ODMR spectroscopy was performed by monitoring the PL signal while sweeping the frequency of microwave excitation delivered by an external loop antenna. 

\indent For all measurements presented below, hBN flakes with an identical thickness of $20$~nm were carefully selected by atomic force microscopy. Using a fixed thickness allows for a direct comparison between samples subjected to different irradiation doses while avoiding thickness-dependent variations of the optical response~\cite{CluaNanoLett2024}. For $20$-nm-thick hBN flakes deposited on a SiO$_2$(85~nm)/Si substrate, we note that the PL signal from \VB center ensembles is maximized due to nanophotonic effects that modify both the absorption and radiation patterns of the emitters~\cite{CluaNanoLett2024}.

\section{Experimental results}
\subsection{Raman spectroscopy}
Before studying the properties of \VB centers, we first analyze the impact of neutron irradiation on the vibrational modes of hBN using Raman scattering spectroscopy. Raman spectra recorded around $1350$~cm$^{-1}$ from monoisotopic hBN crystals irradiated with different neutron doses are shown in Fig.~2(a). The sharp Raman line observed at $1375$~cm$^{-1}$ corresponds to the intralayer $E_{\rm 2g}$ phonon mode, in which $^{10}$B and $^{15}$N atoms vibrate against each other within the hexagonal layer~\cite{PhysRevB.94.155435}. Compared with hBN of natural isotopic composition, the $E_{\rm 2g}$ mode is blue-shifted by $\sim10$~cm$^{-1}$ due to isotopic mass differences~\cite{EdgarAdvMat,vuong2018}. For pristine hBN, the full width at half maximum of the $E_{\rm 2g}$ Raman line is around $3.4$~cm$^{-1}$, reflecting the high crystalline quality of the material. After neutron irradiation, the $E_{\rm 2g}$ linewidth slightly increases to $\sim 4.7$~cm$^{-1}$ and remains nearly constant for crystals S1 to S8, {\it i.e.} for irradiation doses up to $10^{17}$~neutrons/cm$^2$ [see inset in Fig.~2(b)]. For the highest dose (sample~S9), the linewidth finally increases significantly, indicating substantial damage of the hBN lattice. 

\begin{figure}[t]
  \centering
  \includegraphics[width = 8.5cm]{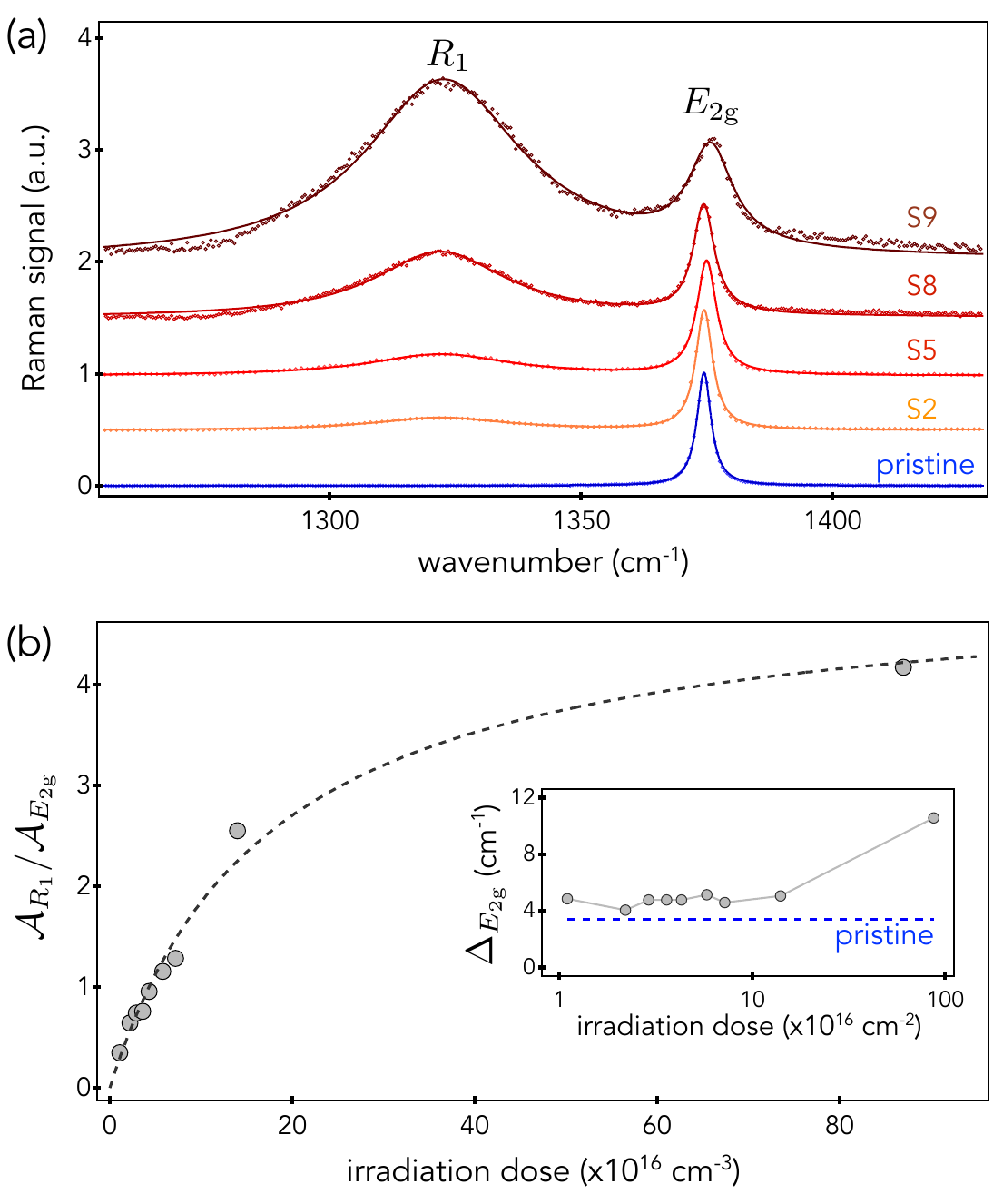}
  \caption{(a) Raman scattering spectra recorded at room temperature from monoisotopic h$^{10}$B$^{15}$N crystals irradiated with different doses of thermal neutrons. Solid lines are fits to the data using a sum of two Lorentzian functions, and the blue curve corresponds to pristine hBN. The data are normalized to the maximum intensity of the $E_{\rm 2g}$ Raman line and vertically shifted for clarity. (b) Evolution of $\mathcal{S}_{R_1}/\mathcal{S}_{E_{\rm 2g}}$ with neutron dose. The black dashed line is a fit using a saturation function. The inset shows the evolution of the full width at half maximum of the $E_{\rm 2g}$ Raman line, $\Delta_{E_{\rm 2g}}$, as a function of neutron irradiation dose.}
\end{figure}

Besides the intralayer $E_{\rm 2g}$ phonon mode, the Raman spectra also feature a broad line at $1322$~cm$^{-1}$, referred to as $R_1$, which has recently been attributed to local vibrational modes associated with \VB centers~\cite{Patra2026}. Our measurements indicate that the $R_1$ line is blue-shifted by $\sim 30$~cm$^{-1}$ compared with hBN of natural isotopic composition~\cite{Patra2026,HeImplant2023,PhysRevMaterials.9.056203}, while its linewidth remains unchanged with increasing neutron irradiation dose. To quantify the dependence of the $R_1$ mode intensity on neutron dose, all Raman spectra were fitted using a sum of two Lorentzian functions. From these fits, the integrated intensity of the $R_1$ line ($\mathcal{S}_{R_1}$) was normalized to that of the $E_{\rm 2g}$ mode ($\mathcal{S}_{E_{\rm 2g}}$). As shown in Fig.~3(b), the ratio $\mathcal{S}_{R_1}/\mathcal{S}_{E_{\rm 2g}}$, which is correlated with the density of \VB centers~\cite{Patra2026}, first increases with irradiation dose and then saturates at high doses. This behavior may be related to the formation of vacancy clusters rather than isolated \VB centers at the highest doses, together with progressive defect-induced amorphization of the hBN lattice. 

\subsection{Optical properties of \VB centers}
\label{opt}
\begin{figure*}[t]
  \centering
  \includegraphics[width = 18cm]{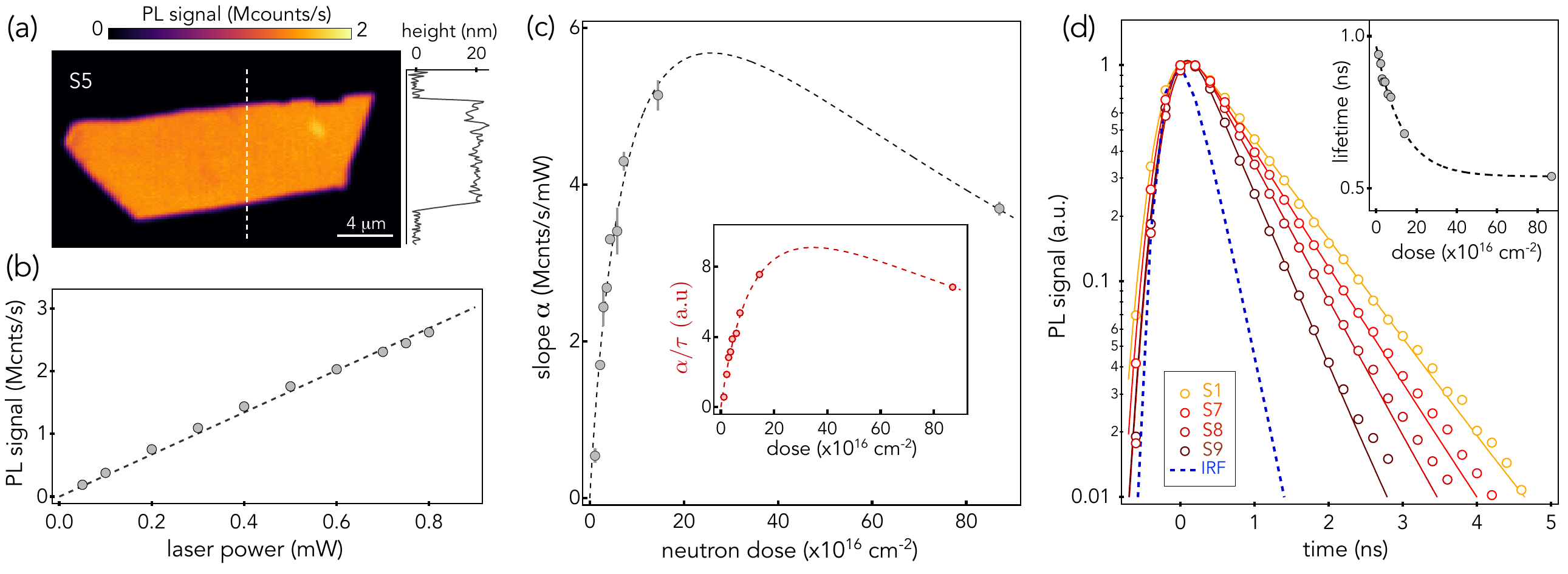}
  \caption{(a) PL raster scan of a 20-nm-thick hBN flake mechanically exfoliated from sample S5. The laser excitation power is set to $500 \ \mu$W. The right panel shows a topographic line-profile recorded across the flake (white dashed line) by atomic force microscopy. (b) PL signal as a function of laser excitation power for the flake shown in (a). The dashed line is a linear fit from which we extract a slope $\alpha=3.36(4)$~Mcounts/s/mW. (c) Parameter $\alpha$ as a function of the neutron irradiation dose. The black dashed line is a guide to the eye. The inset shows the evolution of $\alpha$ normalized by the excited state lifetime $\tau=k_{\rm nr}^{-1}$, as a function of neutron dose. (d) Time-resolved PL decays of V$_{\rm B}^{-}$ centers generated at different neutron doses. The blue dashed line corresponds to the instrument response function (IRF) of the photon counting system measured with $60$-ps laser pulses. Solid lines are fits to the data using a mono-exponential function convolved with the IRF response. The inset shows the evolution of the excited-state lifetime $\tau=k_{\rm nr}^{-1}$ with neutron dose.}
  \label{fig3}
\end{figure*}

We now turn to the optical properties of \VB center ensembles generated at different neutron irradiation doses. Figure~3(a) shows a representative PL raster scan of a 20-nm-thick flake exfoliated from sample~S5, revealing spatially homogeneous emission from \VB centers. In our experiments, the laser excitation power $\mathcal{P}_L$ is always much smaller than the saturation power of the optical transition. The PL count rate thus increases linearly with $\mathcal{P}_L$ [Fig.~3(b)], with a slope $\alpha\propto \rho\times\epsilon $, where $\rho$ is the density of \VB centers and $\epsilon$ denotes the quantum efficiency of their optical transition. The latter is given by $\epsilon=k_{\rm r}/(k_{\rm r}+k_{\rm nr})\approx k_{\rm r}/k_{\rm nr}$, with $k_{\rm r}$ and $k_{\rm nr}$ the radiative and non-radiative decay rates of \VB centers, respectively, which satisfy $k_{\rm r}\ll k_{\rm nr}$~\cite{ivady2020,Reimers2020,Jules2026}. The evolution of $\alpha$ with irradiation dose is plotted in Fig.~3(c): it initially increases with neutron dose, reaches a maximum, and then decreases at the highest dose. 

To gain insight into the origin of PL quenching at high neutron dose, the excited-state lifetime $\tau$ of the \VB center was measured using a pulsed laser source with a pulse duration of $60$~ps. After optical excitation, the relaxation of \VB centers is dominated by non-radiative decay channels involving intersystem crossing to a metastable state, leading to an excited-state lifetime $\tau \approx k_{\rm nr}^{-1} \sim 1$~ns under ambient conditions~\cite{baber2021excited,Clua2024,Whitefield2023}. Time-resolved PL decays recorded from \VB centers generated at different neutron doses reveal a significant reduction in the excited-state lifetime with increasing dose [Fig.~3(d)]. This observation indicates an enhancement of non-radiative decay rates within the optical cycle of V$_{\rm B}^{-}$ centers, likely due to irradiation-induced damage to the hBN lattice. To account for the resulting variations in quantum efficiency, we plot $\alpha k_{\rm nr}$ as a function of neutron dose in the inset of Fig.~3(c). Since $\alpha k_{\rm nr}\propto\rho k_{\rm r}$, this quantity reflects the combined effects of the \VB center density and their radiative decay rate. Interestingly, $\alpha k_{\rm nr}$ slightly decreases at high neutron doses, even though the intensity of the $R_1$ Raman line increases [Fig.~3(b)]. Given the proposed correlation between this Raman intensity and the \VB center density~\cite{Patra2026}, this apparent discrepancy may point to a reduction in the radiative decay rate $k_{\rm r}$ of \VB centers at the highest irradiation doses, which reduces further its quantum efficiency. Overall, these results identify an optimal range of neutron irradiation doses that maximizes the PL brightness of hBN flakes.

\subsection{Spin properties of \VB centers}

We then investigate the impact of neutron irradiation on the electron spin properties of \VB centers. In the absence of an external magnetic field, the Hamiltonian describing the spin triplet ground state ($S=1$) of the \VB center can be written, in frequency units, as
\begin{equation}
\hat{\mathcal{H}_0}=D\hat{S}_z^2 + E(\hat{S}_x^2-\hat{S}_y^2) + \hat{\mathcal{H}}_{\rm h} \  ,
\end{equation}
where $\{\hat{S}_x,\hat{S}_y,\hat{S}_z\}$ are the dimensionless electron spin operators, with the $z$ axis pointing along the $c$-axis of the hBN crystal. The first term describes the spin-spin interaction between the two unpaired electrons of the defect, giving rise to an axial zero-field splitting $D \sim 3.47$~GHz between the spin projections $m_s=0$ and $m_s=\pm1$ along the $z$ axis~\cite{Gottscholl2020}. The parameter $D$ is sensitive to variations of temperature and in-plane strain fields~\cite{GottschollNatCom2021,ACSPhot_Guo2021,ZhaoStrain2026}. The second term results from the coupling of the \VB center to local electric fields produced by surrounding charges~\cite{Gong2023NatCom,Udvarhelyi2023}. This interaction mixes the $m_s=\pm1$ spin sublevels and produces new eigenstates separated by an orthorhombic splitting $2E$, which increases with the density of charges in the hBN crystal. The last term $\hat{\mathcal{H}}_{\rm h}$ describes the hyperfine interaction of the \VB center electron spin with the three nearest $^{15}$N nuclear spins ($I=1/2$), which reads
\begin{equation}
\hat{\mathcal{H}}_{\rm h}=\sum_{i=1}^{3}\mathbf{\hat{S}} \cdot \mathcal{A}^{(i)} \cdot \mathbf{\hat{I}^{(i)}}  \ ,
\label{HamiltoHyp}
\end{equation}
where $\mathcal{A}^{(i)}$ denotes the hyperfine interaction tensor. In the secular approximation and neglecting off-diagonal components of
the hyperfine tensor, this term reduces~to
\begin{equation}
\hat{\mathcal{H}}_{\rm h}=\sum_{i=1}^{3}\mathcal{A}_{zz} \hat{S}_z  \hat{I}^{(i)}_z  \ ,
\label{HamiltoHyp2}
\end{equation}
where $\mathcal{A}_{zz}\sim 64$~MHz for each of the three equivalent nearest-neighbor $^{15}$N nuclei~\cite{Clua2023,Gong2024}. 

\subsubsection{Zero-field splitting parameters}
\begin{figure}[b]
  \centering
  \includegraphics[width = 8.6cm]{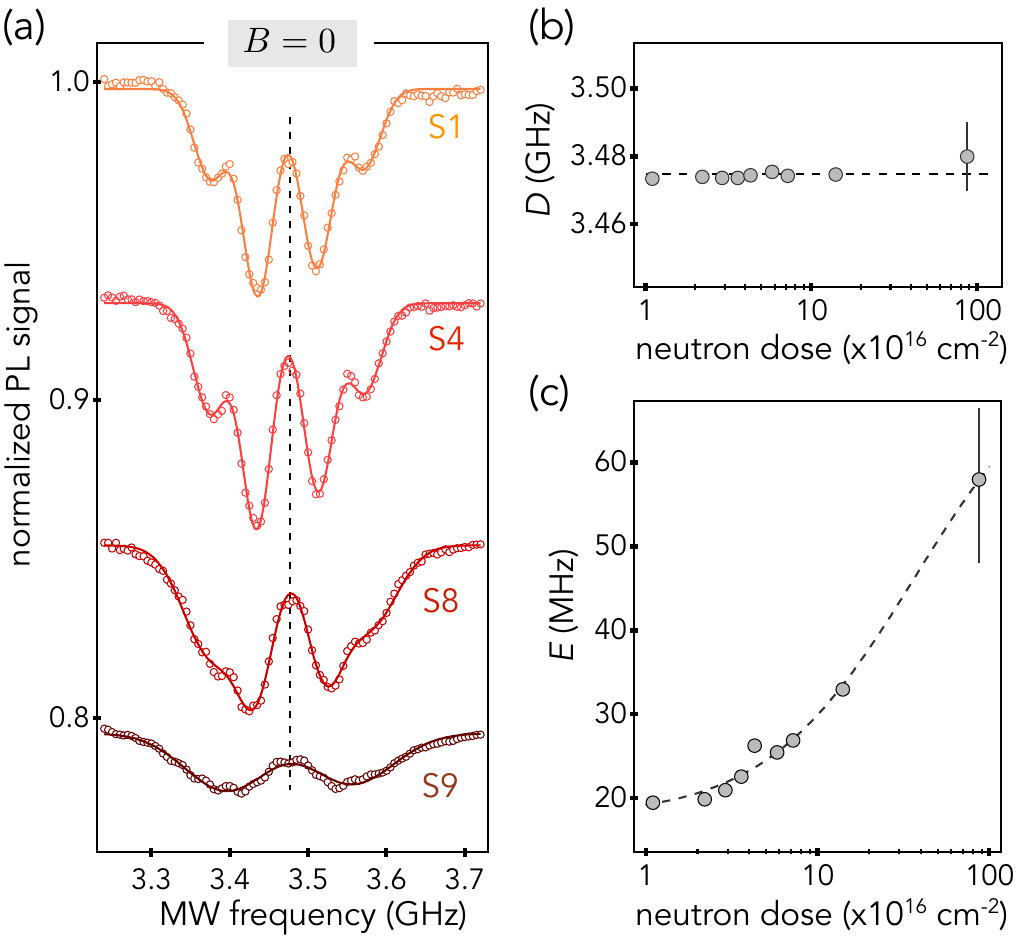}
  \caption{(a) ODMR spectra recorded at zero external magnetic field for different neutron-irradiated h$^{10}$B$^{15}$N samples. Solid lines are fits to the data using a sum of four Gaussian functions, from which the parameters $D$ and $E$ are extracted. The spectra are normalized and vertically shifted for clarity. (b) Zero-field splitting parameter $D$ and (c) orthorhombic splitting $E$ as a function of neutron irradiation dose. The black dashed lines are guides to the eye.
  }
  \label{fig4}
\end{figure}

The spin Hamiltonian $\hat{\mathcal{H}_0}$ gives rise to four magnetic resonances, whose frequencies are given by
\begin{equation}
\nu_{\pm,m_I}=D\pm\sqrt{E^2+(m_I\mathcal{A}_{zz})^2} \ ,
\label{Eigen}
\end{equation}
where $m_{I}=\{\pm 1/2, \pm 3/2\}$ is the total nuclear spin projection of the three equivalent neighboring $^{15} \rm N$ atoms. 

Representative ODMR spectra recorded at zero external magnetic field for hBN crystals irradiated with different neutron doses are shown in Fig.~4(a). Starting from a well-resolved four-line hyperfine structure at low irradiation doses, the magnetic resonances progressively broaden at the highest doses with a significant reduction in ODMR contrast. By fitting the spectra with a sum of four Gaussian functions, we extract the zero-field splitting parameters $\{D,E\}$ using Eq.~(\ref{Eigen}) and a fixed hyperfine coupling strength $\mathcal{A}_{zz}= 64$~MHz. As displayed in Fig.~4(b), no significant variation in $D$ is observed across the different hBN samples, indicating that neutron irradiation does not induce appreciable in-plane strain fields, even at the highest doses. In contrast, the orthorhombic splitting $E$ increases significantly with neutron irradiation dose, reflecting an increase in the local charge density in hBN. Previous studies have proposed a simple microscopic charge model in which the negative charges associated with \VB centers are compensated by concomitant positively charged defects, thereby ensuring charge neutrality of the hBN crystal~\cite{Gong2023NatCom,Udvarhelyi2023}. Within this framework, the local electric field, and hence the $E$ splitting, is directly linked to the density of \VB centers. However, a recent study has suggested that additional sources of local electric fields may become relevant at high irradiation doses, possibly arising from other irradiation-induced charged defects in the hBN lattice~\cite{PhysRevMaterials.9.056203}. Thus, the relationship between the $E$ splitting and the \VB center density may no longer be straightforward, as the measured splitting can also reflect contributions from other charged defects. Despite this caveat, the observed increase in $E$ with neutron dose is consistent with the dose-dependent increase in the intensity of the $R_1$ Raman line discussed above, which has also been proposed as a proxy for the \VB center density~\cite{Patra2026}.

\subsubsection{DC magnetic field sensitivity}

\begin{figure*}[t]
  \centering
  \includegraphics[width = 17.6cm]{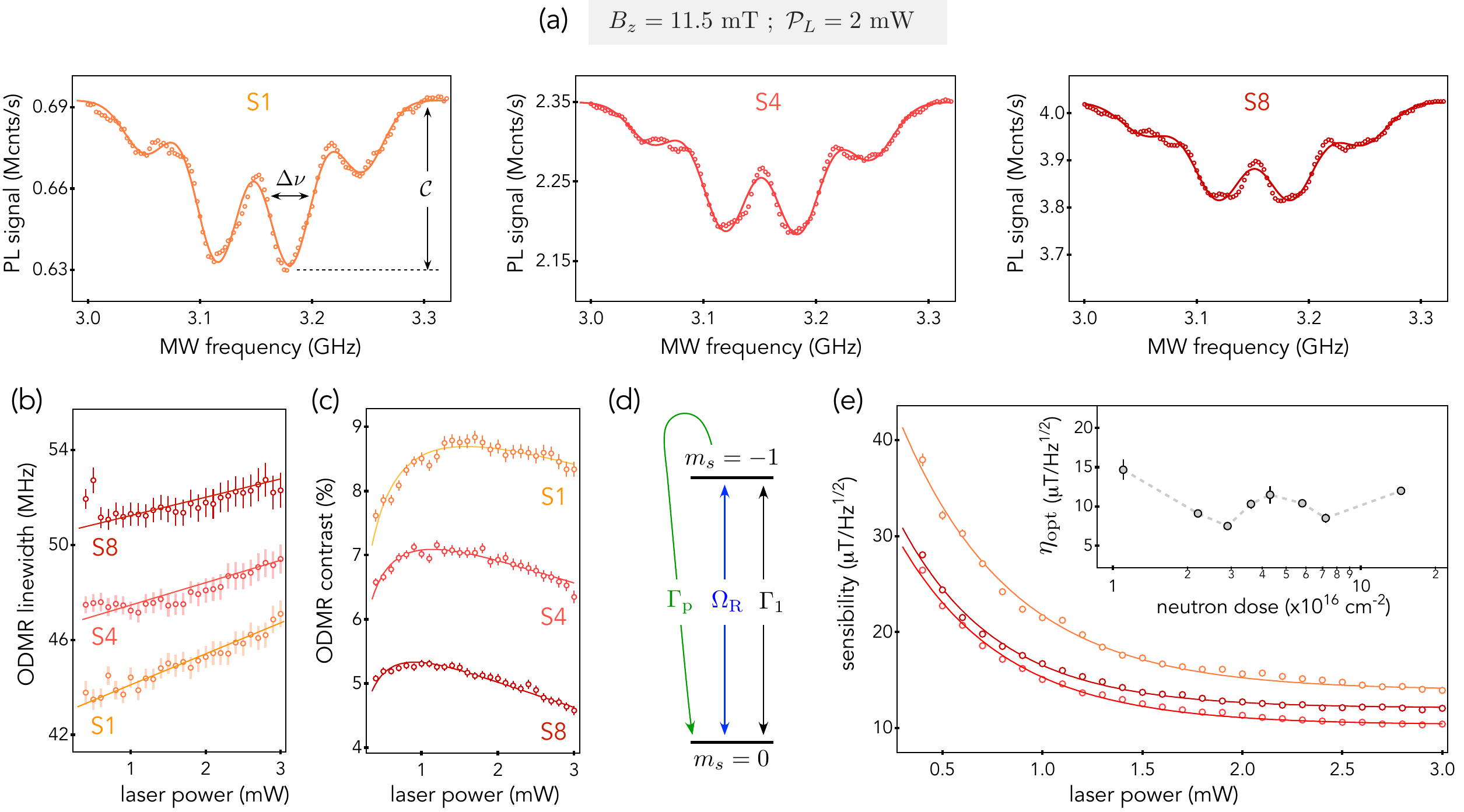}
  \caption{(a) ODMR spectra of the lower frequency components $\nu_{-,m_I}$ recorded from samples S1, S4 and S8 at $B_z=11.5$~mT and $\mathcal{P}_L=2$~mW. Solid lines are fits to the data with a sum of four Gaussian functions. (b,c) ODMR linewidth $\Delta\nu$ and contrast $\mathcal{C}$ of the central hyperfine transitions as a function of optical excitation power $\mathcal{P}_L$. (d) Simplified two-level model of the \VB center. All notations are defined in the main text. (e) Magnetic field sensitivity as a function of $\mathcal{P}_L$ under diffraction-limited laser excitation. The inset shows the optimal magnetic field sensitivity $\eta_{\rm opt}$ as a function of neutron irradiation dose.}
  \label{fig5}
\end{figure*}

We now analyse how the sensitivity of an ensemble of \VB centers to static (DC) magnetic fields evolves with the neutron irradiation dose. To this end, a bias magnetic field $B_z$ was applied along the $z$ axis to lift the degeneracy of the $m_s=\pm1$ spin sublevels via the Zeeman effect, and we focused on the lower-frequency components of the magnetic resonance spectrum, $\nu_{-,m_I}$. When the Zeeman and hyperfine interaction terms dominate over the orthorhombic splitting -- {\it i.e.} for  $\gamma_e B_z + |m_I|\mathcal{A}_{zz} \gg E$ -- these frequency components evolve linearly with the magnetic field as
\begin{equation}
\nu_{-,m_I}=D-\gamma_e B_z + m_I\mathcal{A}_{zz} \ ,
\label{Eigen2}
\end{equation}
where $\gamma_e=28$~GHz/T is the electron spin gyromagnetic ratio. 

Typical ODMR spectra recorded at $B_z=11.5$~mT for 20-nm-thick flakes exfoliated from different neutron-irradiated hBN crystals are shown in Fig.~5(a). As in zero-field measurements, the characteristic four-line hyperfine structure of the \VB centers is well-resolved for all samples except S9, which is therefore excluded from the subsequent analysis of the DC magnetic field sensitivity. By fitting ODMR spectra with a sum of four Gaussian functions, we extract the rate of detected photons $\mathcal{R}$, as well as the contrast $\mathcal{C}$ and the linewidth $\Delta\nu$ of the central hyperfine transitions, which correspond to $m_I=\pm1/2$ [see Fig.~5(a)]. Using these measurements, the shot-noise-limited sensitivity to DC magnetic fields $\eta$ is then given by~\cite{PhysRevB.84.195204} 
\begin{equation}
\eta\approx 0.7 \times \frac{1}{\gamma_e}\times\frac{\Delta\nu}{\mathcal C\sqrt{\mathcal R}} \ .
\label{sensib}
\end{equation}

From ODMR spectra recorded at different optical powers $\mathcal{P}_L$, we first analyze how the parameters $\{\mathcal{R}, \Delta\nu, \mathcal{C}\}$ evolve as a function of $\mathcal{P}_L$ for all hBN samples. As already discussed in Section~\ref{opt}, the detected photon rate $\mathcal{R}$ increases linearly with $\mathcal{P}_L$, though with a reduced slope $\alpha$ owing to partial obstruction of the excitation and collection paths by the external antenna used for microwave excitation. The evolution of $\Delta\nu$ with $\mathcal{P}_L$ is shown in Fig.~5(b) for different hBN samples. Two remarks are in order. First, $\Delta\nu$ only increases by a few percent due to power broadening~\cite{PhysRevB.84.195204}, indicating that the linewidth remains predominantly limited by the spin coherence time ($T_2^\star$) of the \VB center within the investigated optical power range. Second, our results show that $\Delta\nu$ slightly increases with neutron irradiation dose, providing a first indication that spin coherence time decreases with neutron dose.\\
\indent We now turn to the evolution of the ODMR contrast. As shown in Fig.~5(c), the contrast initially increases with~$\mathcal{P}_L$, reaches a maximum, and then decreases at higher optical powers. This behavior, observed for all hBN samples, can be qualitatively understood by modeling the \VB center as a simple closed two-level system corresponding to the ground states with spin projections $m_s=0$ and $m_s=-1$ [Fig.~5(d)]~\cite{PhysRevB.84.195204}. Under continuous laser and microwave excitation, the steady-state spin populations $p_0^{\rm st}$ and $p_{-1}^{\rm st}$ result from the competition between (i) optically-induced spin polarisation into $m_s=0$ with a rate $\Gamma_\text{p}$ increasing with the optical excitation power, (ii) the Rabi frequency $\Omega_{\rm R}$ driving transitions between the two electron spin states, and (iii) the longitudinal spin relaxation rate $\Gamma_1$. In this simplified model, the PL signal of the \VB center ensemble is expressed as
\begin{equation}
\mathcal{R}(\Omega_{\rm R},\mathcal{P}_L)\propto \left[\alpha p_{0}^{\rm st}+\beta p_{-1}^{\rm st}\right]\mathcal{P}_L \ ,
\label{PL}
\end{equation}
where $\alpha>\beta$ to account for the spin-dependent PL response of the \VB center. The ODMR contrast is then defined as
\begin{equation}
\mathcal{C}=\frac{\mathcal{R}(0,\mathcal{P}_L)-\mathcal{R}(\Omega_{\rm R},\mathcal{P}_L)}{\mathcal{R}(0,\mathcal{P}_L)} \ ,
\end{equation}
where $\mathcal{R}(0,\mathcal{P}_L)$ denotes the PL rate in the absence of microwave excitation.  Using a rate equation analysis and assuming $\Omega_R\gg\Gamma_1$, as is always the case in our experiments, the ODMR contrast can finally be expressed as
\begin{equation}
\mathcal{C}=
\frac{(\alpha-\beta)\Omega_{\rm R}\Gamma_\text{p}}
{(\Gamma_\text{p}+2\Omega_{\rm R})\left[\alpha(\Gamma_1+\Gamma_{\rm p})+\beta\Gamma_1\right]}
\ .
\label{contrast}
\end{equation}

In the limit $\Gamma_\text{p}\ll(\Omega_{\rm R},\Gamma_1)$, {\it i.e.} at low optical excitation power, spin polarization into $m_s=0$ is inefficient, leading to a low ODMR contrast that increases linearly with $\Gamma_\text{p}$, and thus with $\mathcal{P}_L$. Conversely, when $\Gamma_\text{p}\gg(\Omega_{\rm R},\Gamma_1)$, the contrast decreases with increasing optical power because optical pumping suppresses microwave-driven population transfer to the $m_s=-1$ state. This simple model thus qualitatively captures the dependence of the ODMR contrast on optical excitation power. Beyond this power-dependent behavior, we also observe a systematic decrease in the ODMR contrast with increasing neutron irradiation dose. This trend suggests a progressive loss in spin selectivity of non-radiative ISC transitions to and from the metastable state during optical cycles~\cite{Clua2024}, which likely results in a reduction of the spin polarisation efficiency.

\indent Using the parameters $\{\mathcal{R}, \Delta\nu, \mathcal{C}\}$ measured at different optical excitation powers, we finally use Eq.~(\ref{sensib}) to determine how the DC magnetic field sensitivity evolves with optical power. As shown in Fig.~5(e), the sensitivity initially improves with $\mathcal{P}_L$ before saturating at an optimal value~$\eta_{\rm opt}$. For the 20-nm-thick hBN flakes exfoliated from samples S1 to S8 -- {\it i.e.} for irradiation doses up to $10^{17}$~neutrons/cm$^2$ -- we obtain similar optimal sensitivities $\eta_{\rm opt}\sim 10 \ \mu$T/$\sqrt{\rm Hz}$ under diffraction-limited laser excitation [see inset in Fig.~5(e)]. This result indicates that the sensitivity enhancement associated with the increased PL rate at higher neutron doses is compensated by a concurrent reduction in ODMR contrast and a slight broadening of the magnetic resonance linewidth. Increasing the neutron irradiation dose up to $10^{17}$~neutrons/cm$^2$ therefore does not lead to a substantial improvement in DC magnetic field sensitivity.  For the highest neutron dose studied in this work (sample S9), the sensitivity is even significantly degraded owing to a strong broadening of the magnetic resonances, which prevents resolution of the hyperfine structure. Finally, we note that the reported sensitivity could be further improved by optimizing the microwave excitation using coplanar waveguides with tailored geometries~\cite{NanoLett2021,CPWNanoLett2025}.\\
\indent It is instructive to compare these sensitivities with those achieved using state-of-the-art wide-field magnetic microscopes based on ensembles of NV centers in diamond. Such systems typically reach DC magnetic field sensitivities in the few-$\mu$T/$\sqrt{\rm Hz}$ range at diffraction-limited spatial resolution, with the sensing NV layer located approximately $10-20$~nm below the diamond surface~\cite{TetienneRevueNV}. An ensemble of \VB centers embedded in a 20-nm-thick hBN flake therefore exhibits a DC magnetic field sensitivity of the same order of magnitude. 
Importantly, unlike diamond-based sensors, the two-dimensional nature of hBN allows \VB centers to be readily integrated with a wide range of material systems, including van der Waals heterostructures and thin-film samples~\cite{Du2021,PhysRevApplied.18.L061002,Tetienne2023,Rita2024,Toeno26,Mu2025,Harvard2025}, providing a more versatile platform for quantum sensing of nanoscale magnetic phenomena. Moreover, the spin-dependent optical properties of \VB centers are preserved down to few-layer-thick hBN flakes~\cite{DurandPRL2023}. Although reducing the hBN thickness would inevitably decrease the sensitivity owing to a reduced number of \VB centers that contribute to the PL signal, the resulting gain in proximity might however provide a substantial net advantage for nanoscale magnetic sensing, since the field generated by a localized source decreases as $1/d^3$ with the sensor-to-sample distance $d$.

\subsubsection{Spin coherence properties}
\label{coh}

\begin{figure*}[t]
  \centering
  \includegraphics[width = 18cm]{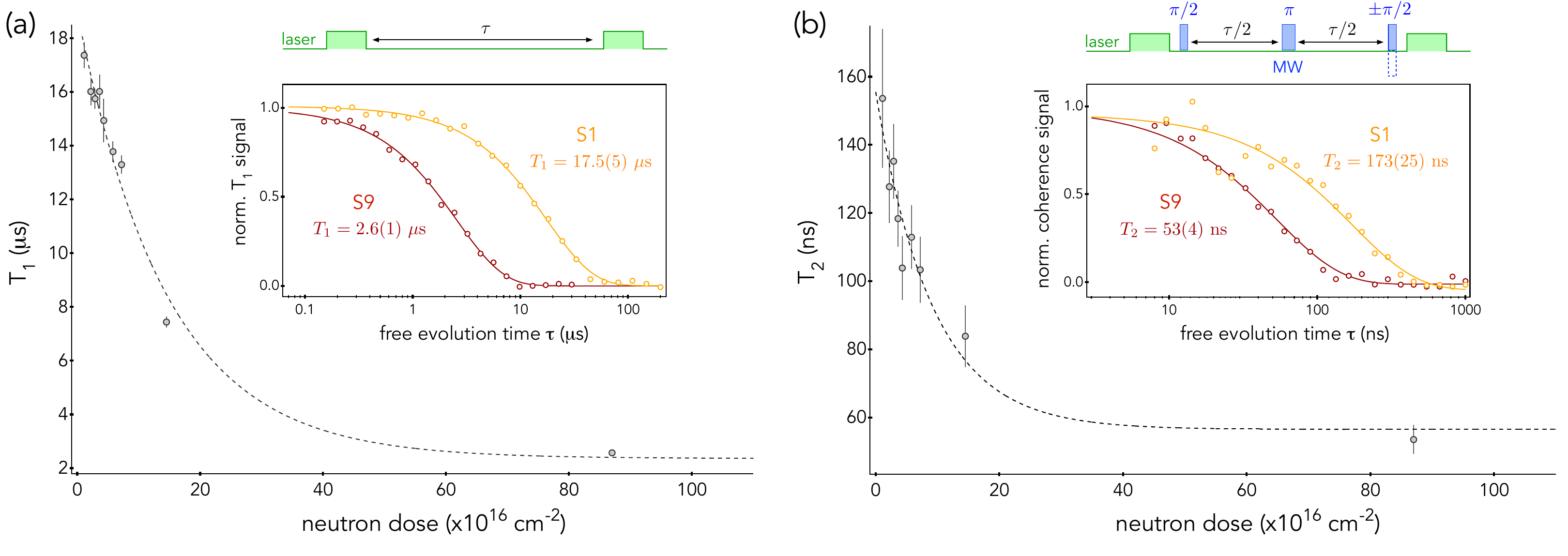}
  \caption{(a) Longitudinal spin relaxation time $T_1$ as a function of the neutron irradiation dose. The inset shows the pulse sequence and the spin relaxation curves measured for samples S1 and S9. The solid lines correspond to mono-exponential fits to the data. (b) Spin coherence time $T_2$ as a function of the neutron irradiation dose. The inset shows the spin-echo sequence and the coherence decay measured for samples S1 and S9. The solid lines correspond to mono-exponential fits to the data. Measurements are performed with a bias magnetic field $B_z\sim10$~mT and a resonant microwave $\pi$ pulse duration of 10~ns.}
\end{figure*}
 
As a final set of experiments, we investigate the impact of the neutron irradiation dose on the spin coherence properties of \VB centers. We begin with measurements of the longitudinal spin relaxation time $T_1$ using the experimental sequence sketched in Fig.~6(a). A first laser pulse is used to polarize the \VB centers into the $m_s=0$ state through optical pumping. Following spin relaxation in the dark during a variable delay time $\tau$, the remaining population in $m_s=0$ is probed by integrating the spin-dependent PL signal recorded at the beginning of a second laser pulse used for readout. The longitudinal spin relaxation time $T_1$ is then extracted by fitting the decay of the integrated PL signal with a mono-exponential function. For the hBN crystal irradiated with the lowest neutron dose (S1), we obtain $T_1=17.5(5)$~$\mu$s, a value commonly observed for \VB centers under ambient conditions when spin relaxation is dominated by spin-phonon interactions~\cite{Gottscholl2021,Lunghi2022,PhysRevB.111.024108}. As the neutron irradiation dose increases, the relaxation time progressively decreases, reaching $T_1=2.6(1)$~$\mu$s for the highest dose (S9). One possible explanation for this behavior could be the enhancement of dipolar interactions between neighboring \VB centers as their density increases. Within a spin-fluctuator picture, dipolar-induced spin flip-flop processes can indeed reduce the longitudinal relaxation time of a dense electron-spin ensemble if a fraction of the defects acts as rapidly fluctuating spins, for instance owing to charge-state hopping~\cite{PhysRevLett.118.093601,PhysRevApplied.20.034050}. However, when such a mechanism dominates spin relaxation, it is expected to produce a stretched-exponential decay with an exponent close to $1/2$ (resp. $1/3$) for a three-dimensional (resp. two-dimensional) dipolar-interacting spin ensemble~\cite{PhysRevLett.118.093601,PhysRevApplied.20.034050}. In contrast, all measured relaxation curves of \VB center ensembles are accurately described by a mono-exponential decay. As an example, for the highest neutron irradiation dose (S9), fitting the relaxation curve with a stretched-exponential function yields an exponent $n=0.99(5)$, confirming the mono-exponential character of the decay [Fig.~6(a)]. This observation suggests that dipolar interactions are not the dominant mechanism responsible for the observed reduction of $T_1$. We tentatively attribute this effect to irradiation-induced modifications of the spin-phonon coupling of the \VB centers, likely arising from the accumulation of lattice damage and the associated perturbation of the local vibrational environment. Future low-temperature measurements, where spin-phonon relaxation is strongly suppressed, will be essential to further elucidate the microscopic origin of this behavior.\\
\indent The spin coherence time $T_2$ was subsequently measured using a spin-echo sequence combined with a differential detection scheme, in which the phase of the final resonant microwave $\tfrac{\pi}{2}$ pulse alternates between $+y$ and $-y$ [Fig.~6(b)]. This protocol efficiently suppresses spin-independent PL contributions, thereby improving the reliability of the extracted spin-echo signal~\cite{Gong2023NatCom}. All measurements were performed with a bias magnetic field $B_z\sim10$~mT. For the hBN flake irradiated with the lowest neutron dose (S1), a mono-exponential fit to the spin-echo decay yields a coherence time of $T_2=173(25)$~ns. This value is close to the theoretically predicted limit set by the interaction of \VB centers with a surrounding bath of $^{15}$N and $^{10}$B nuclear spins under weak magnetic fields~\cite{AFM_Ivady2026,AFM_Seo2026}, indicating that nuclear-spin-induced dephasing likely represents the dominant decoherence mechanism in the low-neutron-dose regime. When the neutron irradiation dose increases, the spin coherence time gradually decreases, reaching $T_2=53(2)$~ns at the highest dose (S9). Such a reduction in $T_2$ is attributed to the emergence of additional decoherence channels associated with fluctuations of a bath of paramagnetic defects including \VB centers themselves and electronic spin $S= 1/2$, which are generated by neutron irradiation~\cite{haykal2021decoherence}. These defects, whose precise microscopic structure remains under debate~\cite{Toledo_2020}, have been identified in neutron-irradiated hBN crystals using electron paramagnetic resonance (EPR)~\cite{Toledo_2020,PhysRevB.98.155203,HUSEYNOV20217218} and cross-relaxation spectroscopy~\cite{haykal2021decoherence}. 

\section{Conclusion}
In conclusion, we have investigated how the optical and spin properties of \VB center ensembles generated by neutron irradiation of monoisotopic h$^{10}$B$^{15}$N crystals evolve over a broad range of neutron doses, based on measurements performed on hBN flakes of identical thickness ($20$~nm). We first showed that the PL brightness of the hBN flakes increases up to a neutron dose of about $10^{17}$~neutrons/cm$^2$, before decreasing at higher doses owing to the combination of a reduction in the quantum efficiency of \VB centers and a saturation of their density. Both effects are likely associated with the formation of extended defects and vacancy clusters at high irradiation doses. We then studied the impact of the irradiation dose on the spin properties of the \VB centers. While the zero-field splitting parameter $D$ of the spin-triplet ground state remains essentially unchanged, the orthorhombic splitting $E$ increases significantly with dose, pointing to a growing contribution from irradiation-induced charged defects to the local electric environment of the \VB centers. By systematically analyzing the characteristics of the ODMR response as a function of neutron dose, we further showed that the DC magnetic field sensitivity of \VB center ensembles remains close to $\eta_{\rm opt}\sim10~\mu$T/$\sqrt{\rm Hz}$ under diffraction-limited laser excitation for doses up to $10^{17}$~neutrons/cm$^2$, before degrading at higher doses. This sensitivity could be further improved by optimizing the microwave excitation using coplanar waveguides to maximize the ODMR contrast~\cite{NanoLett2021,CPWNanoLett2025}. Finally, we found that both the longitudinal spin relaxation time $T_1$ and the spin coherence time $T_2$ of \VB center ensembles are close to their intrinsic limits at the lowest neutron irradiation dose and gradually decrease with dose. These results provide important insights into the development of 2D quantum sensors based on \VB centers in hBN, by identifying neutron irradiation conditions that provide a favorable balance between optical brightness, magnetic field sensitivity, and preservation of spin coherence properties.

\section*{Acknowledgements}
This work was supported by the Agence Nationale de la Recherche through the program ESR/EquipEx+ 2DMAG (grant number ANR-21-ESRE-0025) and the project Qfoil (ANR-23-QUAC-0003), a joined program of the United States National Science Foundation and the Agence Nationale de la Recherche (award
number 2413808 - ANR/NSF/QISE), the Institute for Quantum Technologies in Occitanie and QuantEdu France. Neutron irradiation of the hexagonal boron nitride crystals at the Ohio State University was supported by the U.S. Department of Energy, Office of Nuclear Energy under DOE Idaho Operations Office Contract DE-AC07- 05ID14517 as part of a Nuclear Science User Facilities award (24-4911 DOI: 10.46936/NSUF/60019037) and (23-4683 DOI 10.46936/NSUF/60019038).

\end{document}